\documentstyle [aaspp4,,graphicx]{article}
\begin{document}
\title{
 Constraints on  the Progenitors of Type Ia Supernovae and
Implications for the Cosmological Equation of State}

\bigskip
\noindent
  
\author {Inma Dom\'{\i}nguez}
\affil{Dept. de F\'{\i}sica Te\'orica y del Cosmos, Universidad de Granada,
18071 Granada, Spain\\ inma@ugr.es}
\author {Peter H\"oflich}
\affil{Department of Astronomy, The University of Texas at Austin, 
TX 78712 Austin, USA\\pah@hej1.as.utexas.edu}
\and 
\author {Oscar Straniero}
\affil{Osservatorio Astronomico di Collurania, 
64100 Teramo, Italy\\ straniero@astrte.te.astro.it}
\begin{abstract}
 
Detailed stellar evolution calculations have been performed to 
quantify the influence of the main sequence mass $M_{MS}$ and the metallicity Z of 
 the progenitor on the structure of the exploding 
WD which are thought to be the progenitors of SNe~Ia. 
 In particular, we study the effects of progenitors  on the
brightness decline relation $M(\Delta M_{15})$ which is a corner stone for the use of 
SNe~Ia as cosmological yard-stick. Both the typical $M_{MS}$ and Z can be expected
to change as we go back in time.
 We consider the entire range of potential progenitors 
with 1.5 to 7 $M_\odot$ and metallicities between Z=0.02 to $1\times 10^{-10}$.
 Our study is based on the delayed detonation scenario with specific parameters
 which give a good account of typical light curves and 
spectra. 
 Based on the structures for the WD, detailed model calculations have been performed for the
hydrodynamical explosion, nucleosynthesis and light curves. 
 
 The main sequence mass has been identified as 
 the decisive factor to change the energetics of the 
explosion and, consequently, dominates the variations in the rise-time to decline relation of light 
curves. $M_{MS}$ has little effect on the color index B-V. For similar decline rates $\Delta M_{15}$, 
the flux at maximum brightness relative to the flux on the radioactive tail
decreases systematically with $M_{MS}$ by 
about $0.2^m$. This change goes along with a reduction of the photospheric expansion velocity $v_{ph}$
 by about 2000 km/sec.  A change in the central density of the exploding WD has similar
effects but produces the opposite dependency between the brightness to tail ratio and $v_{ph}$ and,
therefore, can be separated.
 
 The metallicity alters 
 the isotopic composition of the outer layers of the ejecta. 
Selective line blanketing at short wavelengths decreases with Z and changes systematically the 
intrinsic color index B-V by up to $-0.06^m$, and it alters the fluxes in the U band and the UV. 
 The change in B-V is critical if extinction corrections are applied. 
 The offset in the calibration of $M(\Delta M_{15})$ is not monotonic in Z and, in general, remains 
$\leq 0.07^m$. 
 
 We use our results and recent observations to constrain the progenitors, and 
to discuss evolutionary effects of SNe~Ia with redshift. 
 The narrow spread in the fiducial rise-time to decline relation  
 in local SNe~Ia restricts the range of main sequence masses to a factor of 2. 
The upper limit of 1 day for the difference between the local 
and distance sample support the need for a positive cosmological constant. 
 The size of evolutionary effects are small ($\Delta M \approx 0.2^m$) but
are  absolutely critical
 for the reconstruction of the cosmological equation of state.
\end{abstract}
 
\section {Introduction}
 
The last decade has witnessed an explosive growth of high-quality 
data for supernovae both from the space and ground observatories with spectacular results, and new
perspectives for the use of SNe~Ia as cosmological yard sticks and for
 constraining the physics of supernovae.
One of the most important new developments in observational supernova research
was to establish the long-suspected
correlation between the peak brightness of SNe~Ia and their
rate of decline, $M(\Delta M_{15})$,  by means of modern CCD photometry (Phillips 1993).
SNe~Ia have provided new estimates for the value of the Hubble constant ($H_0$)
based on a purely empirical procedure
 (Hamuy et al. 1996ab, Riess, Press \& Kirshner, 1996), and on a
comparison of detailed theoretical models with observations (H\"oflich \& Khokhlov 1996, hereafter HK96;  
Nugent et al. 1997). The values obtained are in good agreement  with one another.
More recently, the routine successful detection of supernovae at 
large redshifts, z (e.g. Perlmutter et al. 1995, 1997; Riess et al. 1998; Garnavich et al. 1998),
has provided an exciting new tool to probe cosmology.  This work has 
provided results that are
consistent with a low matter density in the Universe and, most
intriguing of all, yielded hints for a positive cosmological constant $\Omega_\Lambda $ of
 $\approx 0.7$.
 It is worth noting  that the  differences in the maximum  magnitude
between $\Omega_\Lambda $=0 and 0.7 is  $\approx 0.25^m$ for redshifts between 0.5 to 0.8.
 These results prompted the quest for the nature of the the 'dark' energy, i.e.
cosmological equation of state. Current candidates include a network of topological defects
such as strings, evolving scalar fields (i.e. quintessence), or the classical cosmological constant.
 For a recent review,  see Ostriker \& Steinhardt (2001),  and Perlmutter, Turner \& White (1999b).
 To separate between the candidates by SNe~Ia, the required accuracy has to be better than $0.05$ to
$0.1^m$ (Albrecht \& Weller 2000).
 The results on $\Omega_\Lambda$ and future projects to measure the cosmological
  equation of state  depend  on the empirical $M(\Delta M_{15})$  which is
calibrated locally. This leaves systematic effects as the main source of concern.
 
 Indeed, there is already some evidence that SNe Ia undergo evolution.
It has been argued, that the local SN~Ia  sample covers
 all the possible variations that may come from different 
 progenitors, different explosion mechanisms and environments,
 etc.. For local SNe Ia, the observational and statistical characteristics depend on their environment.
  They occur less often in ellipticals than in spirals, and
 the mean peak brightness is dimmer in ellipticals (Branch et al. 1996; Wang, H\"oflich \& Wheeler 1997;
 Hamuy et al. 2000).  
 In the outer part of spirals the brightness is similar to ellipticals while, in more central regions,
 both intrinsically brighter and dimmer SNe~Ia occur (Wang et al. 1997).
  These dependencies show us that SNe Ia likely depend on the underlying population and
 may undergo  evolution.
 If the evolution realizes then we have to know it and take it into account going back in time.
 Otherwise we can not {\it safely} use a local calibration.
 In principle, more distant sample could come from younger and more metal-poor progenitors, or
the dominant explosion scenario may change.

There is general agreement that SNe~Ia result from some  process of combustion of a
degenerate white dwarf (Hoyle \& Fowler 1960). Within this general picture, three
classes of models have been considered: (1) An explosion of a CO-WD, with mass
close to the Chandrasekhar mass, which accretes mass through Roche-lobe overflow
from an evolved companion  star (Whelan \& Iben 1973). The
explosion is then triggered by compressional heating near the WD center.  (2) An
explosion of a rotating configuration formed from the merging of  two low-mass WDs,
caused by the loss of angular momentum due to gravitational radiation
(Webbink 1984, Iben \& Tutukov 1984, Paczy\'nski 1985).  
 (3) Explosion of a low mass CO-WD
triggered by the detonation of a helium  layer (Nomoto 1980, Woosley et al. 1980, 
Woosley \& Weaver 1986).
Only the first two models appear to be viable.
The third, the sub-Chandrasekhar WD model, has been ruled out on the basis of
predicted light curves and spectra (H\"oflich et al. 1996b, Nugent et al. 1997).

 Delayed detonation (DD) models (Khokhlov 1991, Woosley \& Weaver 1994, Yamaoka et al. 1992)
have been found to reproduce the optical and infrared light curves 
and spectra of 'typical' SNe~Ia reasonably well (H\"oflich 1995, hereafter H95; H\"oflich, Khokhlov \& Wheeler 1995, hereafter
 HKW95; HK96; Fisher et al. 1998;
Nugent et al. 1997; Wheeler et al.  1998; H\"oflich et al. 2000;
 Lentz et al. 2001; Gerardy et al. 2001).
This model assumes that burning starts as subsonic deflagration and then turns to a
supersonic, detonative mode of burning. Due to the one-dimensional nature of the
model, the speed of the subsonic deflagration and the moment of the transition to
a detonation are free parameters.
 The moment of deflagration-to-detonation
transition (DDT) is conveniently parameterized by introducing the transition
density, $\rho_{\rm tr}$, at which DDT happens.
 The amount of $^{56}$Ni, $M_{56Ni}$, depends primarily on
$\rho_{tr}$ (H95; HKW95; Umeda et al. 1999),
 and to a much lesser extent on the assumed value  of the deflagration speed,
initial central density of the WD, and initial chemical composition (ratio of
carbon to oxygen).
 Models with smaller transition density give less nickel and hence both lower peak luminosity
and lower temperatures (HKW95, Umeda   et al. 1999).
 In DDs, almost the entire WD is burned, i.e. the total production of
nuclear energy is almost constant. This and the dominance of $\rho_{tr}$ for the $^{56}Ni$ production
are the basis of why, to first approximation, SNe~Ia appear to be a
one-parameter family.
 The  observed $M(\Delta M_{15})$ can be well understood
as a opacity effect (H\"oflich et al. 1996b),  namely, as a consequence of
the rapidly dropping opacity at low temperatures 
(H\"oflich, Khokhlov \& M\"uller 1993, Khokhlov, M\"uller \&
H\"oflich 1993).
 Less Ni means lower temperature and, consequently,  reduced mean opacities because
the emissivity is shifted from the UV towards longer wavelengths with less line blocking.
A more rapidly decreasing photosphere causes a faster release of the stored energy and, as
 a consequence, steeper declining LCs with decreasing brightness. 
 The DD models thus give a
natural and physically well-motivated origin of the $M(\Delta M_{15})$ 
relation of SNe~Ia within the paradigm of thermonuclear combustion of
Chandrasekhar-mass CO white dwarfs.
 Nonetheless, variations of the other parameters  lead to some
deviation from the $M(\Delta M_{15})$.  E.g. a change of the central density results
in an increased binding energy of the WD and a higher fraction of
electron capture close to the center  which reduce the $^{56}Ni$ production
 (H\"oflich et al. 1996b).
 Because DD-models allow to reproduce the observations, we use this scenario to
 test the influence of the underlying stellar population on the explosion.

We note that detailed analyses of observed spectra and light curves indicate that
mergers and deflagration models such as W7 may contribute to the supernovae population
(H\"oflich \& Khokhlov 1996, Hatano et al. 2000). In particular,
the classical ``deflagration'' model W7 with its  structure similar to DD models
 has been successfully applied to reproduce optical light curves and spectra
(e.g Harkness, 1987).  The evidence against pure deflagration models for the majority of SNeIa includes
 IR-spectra which show signs of explosive carbon burning at high 
expansion velocities (e.g. Wheeler et al. 1998) and recent calculations for 3-D deflagration fronts
by Khokhlov (2001) which predict the presence of unburned and partial burned material down 
to the central regions.  Currently, pure deflagration models may be disfavored
for the majority of SNeIa but, clearly, they cannot be ruled out either.

Previously, H\"oflich, Wheeler \& Thielemann (1998, hereafter HWT98) studied evolutionary effects induced by the  progenitor.
 They calculate differences in the LC and NLTE-spectra
 as a function of parameterized values of the integrated C/O ratio $C/O_{Mch}$ and metallicity of the exploding WD.
This study showed that a change of $C/O_{Mch}$  alters  the energetics of the
explosion which results in  an off-set of the brightness-decline relation. Most prominently,
this effect can be identified by a change in the fiducial  rise-time to decline relation
$t_{Fmax}/t_{F\Delta M_{15}}$. The  offset in $M(\Delta M_{15})$ is given by
$\Delta M_V \approx 0.1 \Delta t$ where 
$\Delta t $ is the dispersion in the rise time  of the 'fiducial' light curve.
 Aldering et al. (2000) showed that $t_{Fmax}/t_{F\Delta M_{15}}$
 are identical within $\Delta t=1d $ for the local and distant sample lending strong
support for the notion that we need a positive $\Omega_\Lambda$.
 A change in the metallicity Z causes a change in the burning conditions at the outer layers
of the WD and it alters the importance of the line blanketing in the blue to the UV.
 Based on detailed calculations, effects of similar order have been found for both the delayed
DD and the deflagration scenario (HWT98, Lentz et al. 2000). Recent studies showed the additional effect
 that Z will influence the final structure of the  progenitor and the resulting
  LCs ( Umeda  et al. 1999, Dom\'\i nguez et al. 2000, H\"oflich
et al. 2000).  However, the former two studies were restricted to the progenitor evolution whereas the
 latter included  the connection between the progenitor and the LC but it was restricted to a 
 progenitor of $M_{MS}= 7M_\odot$
and two metallicities,  Z=0.02 and 0.004.

  A more comprehensive study may be useful to eliminate
potential problems  due to evolution of the progenitors for the determination of the cosmological
equation of state, and it may provide a direct link to the progenitors of SN~Ia.
 In this work we  connect $M_{MS}$ and and the initial metallicity of the WD  to
 the light curves and spectral properties of SNe~Ia for the entire range of
potential progenitors.
 In Section 2 we discuss
 the evolutionary properties of our models. In section 3, the results are presented for the
 explosion, nucleosynthesis, the light curves and spectral properties. In the final, concluding section,
  our model calculations are related to observations, and we discuss constraints  for the progenitors and
 implications for the cosmological equation of state.

\section{The formation of a CO WD}
 
CO white dwarfs are the remnants of the evolution of low and intermediate
mass stars (Becker \& Iben 1980).
Their progenitors are stars less massive than $M_{up}$, which is the lower
stellar mass for which
 a degenerate carbon
ignition occurs after the central helium exhaustion. 
The precise value of  $M_{up}$ depends on the chemical composition (see
Dom\'\i nguez et al., 1999, for a recent
evaluation of this mass limit). It ranges between 6.5 and 8 $M_\odot$. On
the base of updated theoretical stellar 
models of intermediate mass stars, Dom\'\i nguez et al (1999) found final
CO core masses 
in the range 0.55 - 1.04 $M_\odot$, in good agreement with semi-empirical
evaluations of the WD masses
(see e.g. Weidemann 1987). 
  
In this paper, we use CO WD structures obtained by evolving 
models with main sequence masses $M_{MS}$ between
1.5 and 7 $M_\odot$ and metallicities Z between $10^{-10}$ and 0.02.
In the following a label identifies a particular progenitor model, namely
$ApBzCD$ for
 a progenitor with a MS mass of $ A.B ~M_\odot$ and  Z=
$C \times 10^{-D}$.\footnote{z00 stands for Z$=10^{-10}$}
 These models have been obtained  by means of the Frascati Raphson-Newton
Evolutionary Code (FRANEC),
 which solves the full set of equations describing both the physical and
chemical evolution
of a star by assuming hydrostatic and thermal equilibrium and a spherical
geometry (Chieffi \& Straniero 1989,
Chieffi, Limongi \& Straniero, 1998). For a detailed description of the
adopted input physics see Straniero, Chieffi \&
Limongi (1997) and Dom\'\i nguez et al (1999). 

Because we are  interested in the final
chemical structure of a CO WD, let us
recall the main properties and the major uncertainties of the evolutionary
phases during which the CO core forms.

In Table 1 we show some properties of our models.
 In columns 1 to 9 we give: (1) the initial composition (Z and Y),
 (2) the model name,
  (3) the main sequence  mass (in $M_\odot$),
  (4) the mass of the
 CO core at the beginning of the TP phase (in $M_\odot$), (5) the C abundance
(mass fraction) at the center,
  (6) the mass (in $M_\odot$) of the homogeneous carbon-depleted central region,
  (7) the averaged C/O ratio within the final, $\approx$1.37 $M_\odot$, CO
white dwarf after accretion,
 and (8) the $^{56}$Ni mass (in in $M_\odot$) synthesized
during the explosion.
The C and O profiles (mass fraction) of the CO core 
for selected thermally pulsing models are shown in Figs. \ref{z22} to
\ref{CORate}. In particular, 
Fig. \ref{z22} shows the changes induced by a different
initial mass, while Figs. \ref{m3} and \ref{m5}  illustrate the effect of
the metallicity.
  
The internal C and O profiles of a WD are generated in three different
evolutionary phases of the progenitor, namely:
i) the central He burning, 
ii) shell He-burning during the early asymptotic-giant-branch (AGB) phase, and
iii) shell He-burning during the thermally pulsing AGB phase.
As illustrated in the figures, they produce three distinct layers.

The central He-burning  produces the innermost homogeneous layer. This
phase is initially dominated
by the carbon production via the 3$\alpha$ reaction occurring in the center
of a convective core.
Once sufficient $^{12}$C is synthesized, the $^{12}C(\alpha,\gamma)^{16}O$
reactions becomes competitive with 
3$\alpha$. Carbon is partially burned into $^{16}$O. Since the opacity
of a C-O mixture is larger than
 that of a  
He mixture, the extension (in mass) of the convective core increases with
time.  
When the He mass fraction in the convective core is reduced down to
$\approx$0.1,
the He-burning is mainly controlled by $^{12}C(\alpha,\gamma)^{16}O$, and
most of the oxygen in the convective core is synthesized during the late
He-burning. The final
abundances in the innermost region of a WD is strongly dependent on the
duration of the last
$5-10\%$ of the entire He-burning lifetime. In column 7 of table 1,
we report the size (in solar masses) of this innermost homogeneous region,
which corresponds to the maximum extension of the convective core.

The intermediate region of the final C/O structure is characterized by a rising carbon abundance.
It is produced during the early-AGB when the He-burning shell advances in mass
until it
approaches the H-rich envelope. The amount of carbon (oxygen) left behind 
increases (decreases) due to the progressive growth of the temperature in the
shell which favors the 3$\alpha$ reactions
 with respect to
the $\alpha$ capture on $^{12}$C. In addition, the short lifetime does not
allow a substantial conversion of carbon
into oxygen.  
   
Finally,  a thin external layer is built up during the thermally pulsing AGB.
At the beginning of a  thermal pulse, the large energy flux is locally
 produced 
by the 3$\alpha$ reactions. It induces the
 formation of a convective shell that 
 rapidly overlaps the whole inter-shell region. Owing to the large He
reservoir, a huge amount of carbon is produced
 at the base of the convective shell. After few years (10-100 yr depending
on the core mass) the convective shell 
 disappears and a quiescent He-burning takes place. It is during this
longer phase that the 
 $^{12}C(\alpha,\gamma)^{16}O$ reactions convert a certain part of the carbon
 produced during the pulse into oxygen. The C/O ratio left below the
He-rich layer by the He burning-shell depends
 on the rate of the $\alpha$ captures on carbon.   
 Note that the outer 'blip' in the carbon profile is the result of the last
thermal pulse where 
the He has not yet fully depleted. The size, in mass, of this third
layer depends on the duration of the 
thermally pulsing AGB phase. Although it is influenced by the assumed mass
loss rate, it is generally
believed that for M$<$3 M$_\odot$ the CO core cannot increase more than
0.1-0.2 M$_\odot$ during the AGB. 
An even smaller increase of the CO core is expected for larger $M_{MS}$.

The  subsequent phase has been  calculated by accreting H/He rich material
on the resulting CO WD. Note that we have assumed that the progenitor
ejects its H/He-rich envelope prior to the 
onset of the accretion epoch.
The accreted matter has a final C/O ratio of $\approx$1.
When the star reaches a mass close to 1.37 M$_\odot$, ignition occurs close to the
center. $\dot M$ has been adjusted to enforce that
 the thermonuclear runaway occurs at the same central density $\rho_c$ in all models.

\subsection {Dependence on the Main-Sequence Mass}

In Fig. \ref{z22}, the chemical structures of our models are shown as a
function of $M_{MS}$ (for Z=0.02). For low stellar masses,
 the core He-burning happens under lower central temperatures
 (see e.g. Dom\'\i nguez et al. 1999).
This favors the $\alpha$ captures on $^{12}$C, which are in
competition with the 3$\alpha$ resulting in
 a slightly smaller central C/O ratio for low $M_{MS}$.

 However, the size of the region of central He-burning, $M_{cen}$, is increasing with $M_{MS}$.
  In a  star with 7 M$_\odot$, the maximum size of the convective
core is about 0.7 M$_\odot$ while, in the 1.5 M$_\odot$ model, it is only
0.25 M$_\odot$. 
This is the  dominating factor for changes in the mean C/O ratio which, in general,
produces the monotonic relation that $C/O_{Mch}$ decreases with increasing $M_{MS}$ (table 1).

\subsection{ Dependence on  the initial metallicity}

 In Figs. \ref{m3} and \ref{m5}, the chemical structures are given for
various Z for stars with 3 and 5
$M_\odot$, respectively.
The sizes of the innermost homogeneous region is not a monotonic function of Z.
This is due to the peculiarity of very low metallicity intermediate mass
stars (see Chieffi et al. 2001). 
In these stars (Z$\le10^{-10}$), the central hydrogen burning
proceeds via the pp-chain instead of the CNO-cycles, as it happen for
larger metallicity.  
It results in a smaller He-core (Ezer \& Cameron 1971, Tornamb\'e \&
Chieffi 1986). For solar metallicity,
the higher opacities and the steeper temperature gradients produce a
smaller He core (Becker \& Iben 1979,
1980). In all cases the dependence on Z of the  final averaged C/O ratio
(column 7 of table 1) 
is small compared to the effect of main sequence mass.
The variation with Z ranges between 5 and 10$\%$.

\subsection{Uncertainties in the final chemical structure.}

 For the final chemical structure, the most important uncertainty 
is due to the ambiguity in nuclear reaction rate of
$^{12}C(\alpha,\gamma)^{16}O$.  The innermost region is also sensitive to the
treatment of turbulent convection which may affect the duration of the late
central He-burning lifetime and the size of the convective core. 

The rate of the $^{12}C(\alpha,\gamma)^{16}O$ at
astrophysical energies is not well  established (see e.g.
Buchmman 1996, 1997).
The cross
section around the Gamow peak is dominated by ground state transitions
through four different processes:
the E1 amplitudes due to the low-energy tail of the $\rm 1^-$
resonance at $\rm E_{cm}=2.42~ MeV$ and to the subthreshold resonance at
$\rm -45~ keV$, and the E2 amplitudes due to the $\rm 2^+$ subthreshold
resonance at $\rm -245~ keV$ and to the direct capture to the  $\rm ^{16}O$
ground state, both with the corresponding interference terms.
Besides ground state transitions, also cascades, mainly through the E2
direct capture to 
$\rm 6.05~ MeV$ $0^+$ and $\rm 6.92~ MeV$ $2^+$ states, have to be
considered. Obviously a higher rate ($\approx $ factor of 2) of this reaction
reduces the carbon abundance left by both the core and the shell He-burning.
Some indications in favor of an high value for this reaction rate comes
from the rise times to maximum light in
SNe~Ia (HWT98),
and from recent studies of  pulsating WDs (Metcalfe, Winget \&
Charboneeau 2001).

Concerning turbulent convective mixing, it only affects the region of the
WD structure produced during the core
He-burning. The two major uncertainties are related to the possible
existence of breathing 
pulses (see Castellani et al. 1985,
Caputo et al. 1986) and the possible existence of a sizeable convective
core overshoot.
 As recently pointed out by Imbriani et al. (2001), 
since breathing 
pulses increase the late core He-burning lifetime, they significantly
reduce the central C/O. In this regard, they mimic
the effects of a high $^{12}C(\alpha,\gamma)^{16}O$ rate. On the contrary,
convective 
core overshoot does not affect the central C/O, but it may enlarge the size
of the convective core and, in turn, may
produce a larger central, C-depleted region.

In the models presented in this paper,  we have neglected breathing
pulses and convective core overshoot.
All the models, except the 3p0z13LR, have been obtained by means of a high
rate for the  $^{12}C(\alpha,\gamma)^{16}O$
reaction (as given by Caughlan et al., 1985). The 3p0z13LR model has been
obtained by adopting the alternative low rate
presented by Caughlan \&  Fowler (1988).
A comparison between the two models with different
$^{12}C(\alpha,\gamma)^{16}O$ reaction rate is shown in Fig. \ref{CORate}.
Changing the $^{12}C(\alpha,\gamma)^{16}O$ rate from the high to the low
value drastically alters
the chemical profiles of the progenitor. The Carbon
abundance increases by about a factor two. At the time of the explosion, 
the average composition of the WD changes from  oxygen-rich ($C/O_{Mch}=
0.74$) to
carbon-rich ($C/O_{Mch}=1.22$)! The consequences for the LCs are discussed
below. 
Although a variation in the assumed convective mixing
scheme may change the quantitative result of our analysis, the overall 
conclusions and tendencies cannot be significantly altered because
its influence is limited to the innermost part of the
pre-explosive structure.

\section{Explosions, Light Curves and Spectral Properties}

Spherical dynamical explosions and corresponding light curves are
calculated. We consider
 delayed detonation (DD) models, because these
have been found to reproduce the optical and infrared light curves 
and spectra of SNe~Ia reasonably well (H\"oflich 1995b; HKW95; HK96;
Nugent et al. 1997; Wheeler et al. 1998; Lentz et al. 2000; H\"oflich et al. 2000;
Gerardy et al. 2001).
 Model parameters have been chosen which allow
to reproduce light curves and spectra of 'typical' Type Ia Supernovae.

For our set of models, the  differences can be attributed to changes
in the progenitor structure of the CO-WD.
As reference, we use the explosion of a
 progenitor with  5 $M_\odot$ at the main sequence and
solar metallicity (model 5p0z22). At the time of the explosion of the WD,
its central density  is 2.0$\times 10^9$ g/cm$^3$
 and its mass  is close to 1.37$M_\odot$.
 The transition density $\rho_{tr}$ from deflagration to
detonation is chosen to be 2.3$\times 10^7$ g/cm$^3$.

\subsection{Explosion Models}

  Explosion models are  calculated using a one-dimensional
radiation-hydro code (HK96)  that solves the hydrodynamical
equations explicitly by the piecewise parabolic method
(Colella \& Woodward 1984).
 Nuclear burning is taken into account using an extended network of 606
isotopes from n,p to $^{74}Kr$
 (Thielemann, Nomoto \&   Hashimoto 1996 and references therein).
 The propagation of the nuclear burning front is given by the
velocity of sound behind the burning front in the case of a detonation 
wave and in a parameterized form during the deflagration phase calibrated by
 detailed 3-D calculations (e.g.  Khokhlov 1995, 2000; Niemeyer \& Hillebrandt 1995).
  We use the parameterization as described in Dom\'\i nguez \& H\"oflich (2000).
 For a deflagration  front at  distance $r_{burn}$ from the center, we assume
that  the  burning velocity is given by  $v_{\rm burn}=max(v_{t}, v_{l})$,
where $v_{l}$ and  $v_{t}$ are the laminar and turbulent velocities with
$$  v_{t}= 0.15~\sqrt{\alpha_{T} ~ g~L_f},~
 with~
\alpha_T ={(\alpha-1)/( \alpha +1}) ~and ~
\alpha ={\rho^+(r_{\rm burn})/ \rho^-(r_{\rm burn})}. \eqno{[1]} $$
\noindent
Here $\alpha _T$ is the Atwood number, $L_f$ is the characteristic
length scale, and $\rho^+$ and $\rho^-$ are the densities in front of
and behind the burning front, respectively. The quantities
 $\alpha$ and  $L_f$  are directly taken
from the hydro at the location of the burning front
and  we take $L_f=r_{\rm burn(t)}$.
The transition density is treated as a free parameter.
The description of the
deflagration front does not significantly  influence
 the final structure of the explosion (Dom\'\i nguez \& H\"oflich 2000).  The
 total $^{56}$Ni production is governed by
 the pre-expansion of the WD
and, consequently, is determined by the transition density $\rho_{tr}$,
at which the burning front switches from
the deflagration to the detonation mode (H95).
 From the physical point of view, $\rho_{tr}$ should be regarded as a convenient way 
to adjust the amount of material burned during the deflagration phase.
The  value $\rho_{tr}$ can be adjusted to produce a given amount of $^{56}Ni$.
  This code 
    includes the solution of the frequency-averaged radiation transport
implicitly via moment equations, expansion opacities, and a detailed
equation of state (see sect. 3.2).
 As expected from previous studies (see introduction),
the overall density, velocity and chemical structures are found to be rather insensitive
to the progenitor, including the production of elements. 

 Although explosions and light curves have been calculated for the entire set of stellar cores,
we will concentrate our detailed discussion on the extreme cases and the reference model.
 Results for intermediate models can be understood accordingly and interpolated using the quantities
given in table 1.

The final density, velocity and chemical  structures and detailed production of isotopes
 are shown in Figs. \ref{rho},   \ref{chem} and  \ref{isotope} for the
 extreme cases in metallicity (Z=0.02 and Z=10$^{-10}$, 5 M$_\odot$) and 
the extremes in $M_{MS}$ (1.5 M$_\odot$ to 7 M$_\odot$, Z=0.02).
 Between 0.511 to 0.589 $M_\odot $ of $^{56}Ni$ are produced (table 1). The production of individual
isotopes varies only by about 10 \%  (Fig. \ref{isotope}). For the reference model $5p0z22$, the
 final element abundances are given in Table 2.
 Variations in the final density and velocity structure are correspondingly small (Fig. \ref{rho}).

 In delayed detonations, almost the entire WD is burned. The total release of nuclear energy depends
mainly on the fuel, i.e. on the integrated C/O ratio $C/O_{Mch}$ (HWT98). However,
  as usual for delayed detonation models, the deflagration phase is key for our understanding of the final
results.
 During the deflagration phase, about $0.33 M_\odot$ of fuel are burned in our models (Figs. \ref{rho} \& \ref{chem}).
 In all explosions but the progenitors with $M_{MS}=1.5\& 3 M_\odot$ with Z=0.02 and $MS=1.5 M_\odot$ with Z=0.001,
  the deflagration front
will propagate in the carbon-depleted layers.
The amount of total energy produced during the deflagration phase and the
binding energy of the progenitor determines the pre-expansion of the outer layers 
and, consequently, the overall chemical structure.
 The binding energy of the WD is dominated by the central density $\rho_c$ at the time of the 
 explosion.
Note that the
 C/O ratio has little influence on the structure of the WD because the pressure is dominated by
degenerate electrons and the electron to nucleon ratio $Y_e$ is identical for $^{12}C$ and $^{12}O$.
  The total energy production during
the deflagration phase is governed by $\rho_{tr}$ and by the nuclear energy release per gram, i.e.
the composition. Both $\rho_{c}$ and $\rho_{tr}$ have been kept the same in all  models.
 Variations can be understood by the change of the mean C/O ratio and the mass $M_{Cen}$ 
 of the central, carbon
depleted region. The latter influences the temperature and, consequently, the laminar speed and the
Atwood number (eq. 1). At the central layers, all the material is burned up to iron-group elements (Fig. \ref{chem}).
  Some additional variation in the total $^{56}Ni$-mass is caused by the C/O ratio of the matter burned during the detonation phase.
For all models, the transition between $^{56}Ni$- and Si-rich layers is  between 0.58 and 0.98 $M_\odot$. For $M_{MS}=7 M_\odot$
and, to a much lesser extend, for $M_{MS}=5 M_\odot$,
this transition region overlaps with layers of lower C-depleted (Fig. 1). 
 From the nuclear physics, the transition between complete and incomplete Si burning occurs in a narrow temperature range around
$5 \times 10^9 K$. A locally lower C/O ratio results in lower burning temperatures. Consequently, the  $^{56} Ni$/Si boundary   is
shifted inwards and $M_{56Ni}$ is reduced.
 Overall, differences between the models remain small because the nuclear energy production for
C- and Oxygen burning differs only by  $\approx 10 \% $.

\subsubsection { Dependence on the Main-Sequence Mass}
 An increasing $M_{MS}$ for a given metallicity leads to changes in
 $C_{cen}$ and $C/O_{Mch}$  by 33\% and -25\%, respectively (table 1).
Increasing $M_{MS}$ from 1.5 to 7.0 $M_\odot$ alters
the expansion velocity of a given mass element (Fig. \ref{rho}, right panel) and it results in  a shift of
the chemical interfaces between complete, incomplete Si and explosive C burning by
 $\approx 1500 km/sec$ (Fig. \ref{chem}).

\subsubsection{ Dependence on  the initial metallicity Z}
 If we decrease the  metallicity from solar ($Z=0.02$) to $Z=10^{-10}$ for stars with $5 M_\odot$,
 $C_{cen}$ and $C/O_{Mch}$ change  by as little as +10 \% and -7 \%, respectively. This means that the density and
velocity structure of the chemical structures are virtually indistinguishable (Fig. \ref{rho}).
 For stars with $M_{MS}$  $\leq 3 M_\odot$, variations in $C_{cen}$ increase up to 30 \% but, still,
variations in $C/O_{Mch}$  remain at a level of 10 \%.  The overall energetics, density and velocity
structure remain mostly unchanged but the pre-expansion and, consequently, the chemical interfaces between
different regimes of burning change by $\approx 200 km/sec$.
 For all $M_{MS}$, the most noticeable difference is the increasing $^{54}Fe$ production with Z in the layers of incomplete
Si burning which changes  the spectra in the blue,  and in the UV (see HWT98 and below).

\subsubsection{Influence of the $^{12}C(\alpha , \gamma )^{16}O$ rate}
As mentioned  in  Sect. 2.3, there is some indirect  evidence for a high cross section
of this key reaction but a low rate cannot be ruled out either. The consequences of a low rate are strong.
E.g.,  for a progenitor with  $M_{MS}=3.0 M_\odot$ and $Z=0.001$,
 the low rate suggested by Caughlan \& Fowler (1988)  will increase  $C_{cen}$ and $C/O_{Mch}$
from 0.26 to 0.51 and 0.74 to 1.22 when compared to our favorite rate (Caughlan et al. 1985).
The explosion becomes more energetic by about 20 \% and the deflagration front propagates faster.
 The result is an increase of the $^{56}Ni $ production by about 10 \% and a shift in the chemical
interfaces by about +2500 km/sec.

\subsection{Light Curves and Spectral Properties}

Based on the explosion models, the subsequent expansion, bolometric and monochromatic light curves 
are calculated (H\"oflich et al.  1998, and references therein). The  LC-code 
is the same used for the explosion, except that $\gamma$ ray 
 transport is included via a Monte Carlo scheme and nuclear burning is neglected. In order to allow a more consistent treatment of
the expansion, we solve the time dependent, frequency averaged radiation moment equations.
 The frequency-averaged variable Eddington factors
and  mean opacities are calculated by solving the frequency-dependent
transport equations.
About one thousand frequencies (in one hundred frequency groups) and
about nine hundred depth points are used.  
 At each 
time step, we use T(r) to determine the Eddington factors and mean opacities by solving the frequency 
 dependent radiation transport equation in a comoving frame and integrate to obtain the frequency 
 averaged quantities. 
The averaged opacities have been calculated assuming local thermodynamics equilibrium (LTE).
 Both, the monochromatic and mean opacities are calculated in the narrow line limit.
 Scattering, photon redistribution and thermalization terms used 
 in the light curve opacity calculations are taken
into account. In previous works, the photon  redistribution and thermalization terms have been
calibrated for a sample of spectra using the formalism of the equivalent two level approach
(H95). Here, for increased consistency,  we use the same  equations
 and atomic models but
solve the rate equations simultaneously with the light curves calculation at about every 100$^{th}$
 time step, on the expense of some simplifications in the NLTE-part compared to H95.
  For the opacities, we use the narrow line limit
and for the radiation fields, we use the solution  
of the monochromatic radiation transport using $\approx 1000 $
frequency groups. Both the old and new approach
 are about equivalent in accuracy with consistent
results. Most noticeable, now,  B-V is  bluer by about $-0.03^m$.
 
 In the following discussion is based on the same
set of models used in the previous section. In Figs. \ref{lc} and \ref{dep}, we show the B and V light curves and some
quantities at maximum light.
 Overall the  different phases of light curves can be understood in the usual way including
the bump at about day 35 which can be attributed to the  change in the opacities between
the layers of complete and incomplete burning (Dom\'\i nguez 1991, 1994).
For the reference model 5p0z22, a maximum brightness $M_V$ of $-19.20 ^m$ is reached at about 18.25 days after
the explosion. The color index $B-V $ is $-0.02^m$.

As discussed in the introduction,
the amount of $^{56}$Ni, its distribution and the expansion rate of the envelope are the dominant
factors which determine the absolute magnitude at maximum and the light curve shape.
 With all model parameters fixed but the progenitor mass and the metallicity, the differences of the
light curves can be understood based on the previous discussion of the explosion models.

\subsubsection { Dependence on the Main-Sequence Mass}
 By increasing $M_{MS}$ from 1.5 to 7.0 $M_\odot$ for Z=0.02,  both $M_B$ and $M_V$  decrease by
 $ \approx 0.15^m$ consistent with a change in the $^{56}Ni$ mass by 14 \% (Fig. \ref{lc}, upper panel, and Fig. \ref{dep}).
  The similarity in the density and velocity structures
  produces almost identical
 conditions at the photosphere. Thus,
B-V is insensitive to a change in $M_{MS}$
 ($\Delta (B-V)[model-5p0z22] \leq 0.01^m$). 
  Relative to the reference model,
the rise times vary between
-0.5 d (1p5z22) to +1.2d (7p0z22). The decline rate  $\Delta M_{15}$ is hardly affected.
 A change in $M_{MS}$ will result in an offset/dispersion in $M(\Delta M_{15})$ by up to $0.15^m$.
 Interestingly, the fluxes on the radioactive tail are much more similar than could be expected from the
spread in the $^{56}Ni$ masses by 14 \% (Fig. \ref{lc}).
 The change in  $M_{56Ni}$ is almost compensated
 by the differences in the energy deposition of $\gamma $-rays from the radioactive decay.
 In Fig. \ref{gam}, the escape probability for hard radiation is shown as a function of time.
 A significant fraction of $\gamma $ photons is thermalized up to about 150 to 200 days.
 The actual value of thermalization depends on the expansion rate which is decreasing with mass (see above).
 E.g. the fraction of thermalized $\gamma $ photons for the models 1p5z22, 5p0z22 and 7p0z22 are 24.2 \%, 25\% and 27 \%, respectively.
The increase in the efficiency for the thermalization  amounts to 11 \% over the mass range and almost compensate the
decrease in the $^{56}Ni$ mass. 
  Note that the  ratio between  maximum and tail brightness is decreasing  with $M_V$. This effect is opposite to the
observed $M(\Delta M_{15})$ relation (e.g. Hamuy et al. 1996). If realized in nature,
 a wide range in $M_{MS}$ would increase the dispersion
in $\delta M(\Delta M_{15})$ by about $0.15^m$. The presence of this effect would reveal itself by an additional change in the expansion velocity
measured by the Doppler shift of lines. E.g. at maximum light, weak lines would indicate an expansion velocity at the photosphere which 
is  smaller by $\approx  2000 km/sec$ if we compare model  7p0z22 vs. 1p5z22.
 The discussion above applies to all metallicities
 because  $C_{cen}$ and $C/O_{Mch}$ vary in a similar range.

\subsubsection{ Dependence on  the initial metallicity Z}
 For progenitors with  $M_{MS} = 5 M_\odot$, a change in the metallicity has little effects on the energetics
 and, consequently, on the light curves
(Fig. \ref{lc}, middle panel). The most important effect is the systematic decline of B-V by $\approx 0.05 ^m$ when Z is changed from
0.02     to $10^{-10}$.  This effect can be attributed to a change of the line blending by Fe in the decoupling region of photons, i.e. the
atmosphere (HWT98, Lentz et al. 2000).
  In B and V at maximum light,  opacities are
 dominated by electron scattering (HKM93) but iron lines are more important in B compared to V.
   Consequently, a lower metallicity results in increase of the flux ratio between B and V.
  In the U-band  and the UV, the opacities are dominated by lines.  A change in the line blending will cause both a change of
the radius  of the flux formation and the specific flux. Therefore, a decrease in Z may result in either an in- or decrease of
the monochromatic flux depending on the density structure.
 These findings apply to all MS-masses in our sample. To some extend, the  exception are the models with $M_{MS}=3M_\odot$ for which
the central carbon concentration varies with metallicity and produces a change in  $M_{56Ni}$ by 3 \% and
 a corresponding change in  $M_B$ and $M_V$.

\subsubsection{Influence of the $^{12}C(\alpha , \gamma )^{16}O$ rate}
A low nuclear rate $^{12}C(\alpha, \gamma)^{16}O$ increases the explosion energy compared to our standard 
rate (3p0z13LR vs. 3p0z13, see Fig. \ref{lc}, lower panel).
The rise time in V  is reduced by 2.7 days (15.3d for 3p0z13LR vs. 18.0d for 3p0z13).
 The enhanced escape probability for $\gamma $-rays (Fig. \ref{gam})
explains the remaining differences including the increased maximum brightness to tail ratio and the moderate increase
of $M_V$ and $M_B$.
  These results are consistent with previous findings which identified the importance of the C/O ratio
for the change in the rise time of 'typical' SNe~Ia (HWT98).

\section{Final Discussion, Observational Constraints and Conclusions}

 Using a delayed detonation model and realistic structures for the exploding white dwarf,
we have studied the influence of the progenitor star on the light curves and spectral properties 
of Type Ia Supernovae.

\noindent
{\bf Stellar models:} We considered stars between 1.5 to 7 $M_\odot$ and metallicities  between
$Z=0.02$ (solar) to $Z = 10^{-10}$ which covers the full range of potential progenitors.
The progenitor structures are based on detailed calculations for the stellar evolution
starting at the pre-main sequence up to the thermal  pulses
when most of the stellar envelope is ejected and a white dwarf is formed with a mass between 0.5 and 1.0 $M_\odot $.
 Its size increases with  $M_{MS}$ and, to a lesser extend, changes with
the metallicity.
 The subsequent accretion and  burning at the surface of the WD let it grow to $M_{Ch}$.
 As a final chemical structure, the WD shows a central region of reduced C abundance between
0.21 to 0.32
originating from the convective He-burning,
a  layer of increased C abundance from  the
He-shell burning, and a layer originating for the accretion phase. The mean C/O ratio
 decreases by about 30 \% over the entire mass range.
 The sensitivity on the metallicity is much weaker ($\leq 10 \%$), and not monotonic.

\noindent
{\bf Supernovae:}
 Our study of SNe~Ia is based on 
  delayed detonation models because
they have been found to reproduce the monochromatic light curves and
and spectra of SNe~Ia reasonably well including the brightness decline relation
  $M(\Delta M_{15})$.
 Deviation from a perfect relation are due to
variations in the central density, properties of the deflagration front, and the progenitor structure.
All parameters but the progenitors have been  fixed to produce
  LCs and spectra typical for 'normal' SNe~Ia. In this work,
rise times to maximum light are between 17.7 to 19.4 days, $M_V = -19.25^m$ to $-19.11^m$, and
 $ B-V = +0.02^m $ to $ -0.07^m$.
 Differences between the models and light 
 curves remain small because the nuclear energy production by burning
Carbon and Oxygen to iron-group elements differs by as little as $\approx 10 \% $.

 The change of $M_{MS}$  is the decisive factor to change the energetics.
The $^{56} Ni$ production varies by about 14 \% and the velocities of the various chemical layers differ
by up to 1500 km/sec. A change in the metallicity hardly affects the overall structure of the progenitor.
As already discussed in detail in HWT98, its main effect is a change in the
production of $^{54}Fe $ in the outer layers of incomplete Si burning.

As one of the main results of our study, we find  that variations in $M_{MS}$ change 
the shape of the LCs but hardly affects B-V whereas a change in Z affects B-V.

  $M_{MS}$ alters the $M(\Delta M_{15})$ relation which may be offset
by up to $0.2^m$. In addition, $M_{MS}$ changes the 
 flux ratio between maximum light and the radioactive tail, and it alters
 the photospheric expansion velocities $v_{ph} $  measured by the Doppler shift of lines.
 E.g. a change in  $m_V(t_{max}) - m_V(t_{max}+40d)$ by $0.2^m$ is coupled to a decrease in
 $ v_{ph}$ at maximum light by $\approx -2000 km/sec$.
 Note that a change in the central density $\rho_c$ of the WD has a similar effect on
$m_V(t_{max}) - m_V(t_{max}+40d)$  but with the opposite sign for $\Delta v_{ph}$
(H\"oflich, 2001).  In principle, this allows 
 to decide whether differences in $m_V(t_{max}) - m_V(t_{max}+40d)$
between SN with similar $M(\Delta M_{15})$ are related to a change in the progenitor or the central density
at the thermonuclear runaway which is sensitive to the accretion rate.

 In contrast to $M_{MS}$, the metallicity Z hardly changes the light curve shapes
 ($\delta M(\Delta M_{15})  \leq 0.06^m$). It
 alters the line blocking by iron group elements at the photosphere mainly in
the UV, U and B but hardly in V (HWT98). In the models presented here, B-V becomes
systematically bluer with decreasing Z (up to  $\approx 0.07^m$).
 Because B-V is the basic color index used to correct for  interstellar extinction, the metallicity
effect can systematically alter the estimates for the absolute brightness by up to $0.2^m$.

\noindent
{\bf $\bf ^{12}C(\alpha, \gamma)^{16}O$:}
 At the example of a progenitor with $M_{MS}=3 M_\odot$ and Z=0.001, we have tested the influence of  the
low nuclear rate $^{12}C(\alpha, \gamma)^{16}O$ on the outcome.
Using the  lower rate suggested by  Caughlan \& Fowler (1988) instead  Caughlan et al. (1985)
 results in more energetic explosions  because $C/O_{Mch}$
 increases by a factor of $\approx  2$.
The rise times to maximum light are 15.3d instead of 18.0d.
 From detailed observations of nearby supernovae, Riess et al. (1999) find the following relation
between the rise time $t_V$ and the maximum brightness
$$ t_V = 19.4 \pm 0.2d + (0.80  \pm 0.05 d ) \times ( M_V + 19.45^m )/ 0.1 ^m. \eqno{[2]}$$
The theoretical LCs
peak at $M_V=-19.21^m$ and $M_V=-19.30^m$ for $3p0z13$ and $3p0z13LR$, respectively.
 From the empirical fit of Riess,  we would expect a rise time between 17.6 and 18.4 days
which favors the high rate $^{12}C(\alpha, \gamma)^{16}O$  of Caughlan et al. (1985). 
 Note that the uncertainties in absolute values for the rise times are $\approx $ 1 to 2 days 
(HWT98). 
 
\noindent
{\bf Constraints on the progenitor:}
 Empirically, the $M(\Delta M_{15})$  has been well established with a rather small statistical
error $\sigma$ ($0.12^m$: Riess et al. 1996;
 $0.16^m$: Schmidt et al. (1998); $0.14^m:$ Phillips 1999; $0.16^m:$ Riess et al. 1999;
$0.17^m:$ Perlmutter et al. 1999a). From theoretical models, a spread of 0.3 to $0.5 ^m$
can be expected (H\"oflich et al. 1996).
  This may imply a correlation between free model parameters, namely the properties 
of the progenitors, the central density or the transition density $\rho_{tr}$.
 In this study, we find a spread in $M(\Delta M_{15})$ of about $0.2^m$ for progenitors with $M_{MS}$
 between $1.5$ to $7.0 M_\odot$. This may suggest a more narrow range in $M_{MS}$ for
realistic progenitors. From the fiducial rise time, progenitors with $M_{MS}$ $\geq 3 M_\odot $ are favored.
This number should be regarded as a hint because uncertainties in the LC models.
 Another constraint can be obtained from the observed spread in the rise-time to decline relation.
Riess et al. (1999) find a spread in $t_V$  of  $\pm 0.4$ days whereas our models 
 show a spread of $\approx $ 1.7 days for $ 1.5 \leq M_{MS} \leq 7 M_\odot$.
 To be consistent with the observations, the range of main sequence masses
has to be reduced by a factor of two. This range is an upper limit because additional variations
in the population of SNeIa such as explosion scenarios or the central density of the WD at the
time of ignition will likely result in lower correlations.

\noindent
{\bf The cosmological equation of state:} In the following, we want to discuss our results in context of
SNe~Ia as probes for cosmology and for the determination of the cosmological equation of state. We limit the
discussion to the effects due to different progenitors. For a discussion of other systematic effects
 such as grey dust, gravitational lensing,
  the influence of a change in the importance of  different possible scenarios
(e.g merger vs. $M_{Ch}$ models)
  etc. we want to refer to the growing literature in this field (e.g. Schmidt et al. 1998,  HWT98,
   Perlmutter et al. 1999a).

 Recently, there has been strong evidence for a positive cosmological constant (e.g. Perlmutter et al.
1999a, Riess et al. 1999). This evidence is based on observations that SNe~Ia in the redshift range
between 0.5 to 1.2 appear to be dimmer by about $0.25^m$ for redshifts between 0.5 to 0.8
 which is comparable to the variations produced
by different progenitors. However, both the internal spread in $M(\Delta M_{15})$
 (see above) and the similarity in 
 $M(\Delta M_{15})$ between the local
SNe~Ia and the high-z sample ($\Delta t \leq 1 d$, Aldering et al. 2000) limit the likely range
of models and, consequently, evolutionary effects to $\leq 0.1^m$ up to  redshifts of $ 1$.
 In addition, we do not expect a drastic change in the metallicity between local and supernovae
at $z\leq 1$.
Taking the linear dependence of B-V on the metallicity (Fig. \ref{dep}) and realistic ranges for z,
also reddening of 'non-grey' dust will not change the conclusion on the need for a
positive cosmological constant.

 As discussed in the introduction, the quest for the nature of the 'dark' energy is one of the
central questions to be addressed in future (e.g. White 1998, Perlmutter et al. 1999b,
 Albrecht \& Weller 2000, Ostriker \& Steinhardt 2001).
 For $z\ge 1$, we can expect both very low 
 metallicities and a significant change  of the typical $M_{MS}$.
From this study, systematic effects due to different progenitors
 are limited  to $\approx 0.2^m$. Therefore,
without further corrections for the progenitor evolution, some of the alternatives for the nature of the
'dark energy' may  be distinguished without correction for evolution.
However, for a more detailed analysis, an accuracy of about $0.05^m$ (Weller \& Albrecht 2001)
is required. In this paper, we have shown how a combination of spectral and LC data or different
characteristics of the LC can help to achieve this goal.

\noindent
{\bf Limitations:} Finally, we also want to mention the limitations of this study.
Qualitatively, our results on the Z dependence agree with a
previous study (H\"oflich et al. 2000) which was based on a progenitor  with $M_{MS} =7~M_\odot$ calculated
by Nomoto's group (Umeda et al. 1999).
 The relation  between  $C/O_{Mch}$ and the offset in the $M(\Delta M_{15})$ relation
 has been confirmed. However, the influence of Z on $C/O_{Mch}$ was found to be about
twice as large, and the central C concentrations  are systematically
larger. The differences point towards a general problem. The details of the central structure and evolution
in  the convective
He burning core  depend sensitively on the treatment of convection,
semi-convection, overshooting, and breathing pulses  (Lattanzio 1991, Schaller et al. 1991,
Bressan et al. 1993, Vassiliadis \& Wood 1993). For a detailed discussion, see Dom\'\i nguez et al.
(1999). In particular, the central C concentration may vary between 0.1 and 0.5.  We want to note,
 that our value is consistent with direct measurements of the central C/O ratio found by the analysis of pulsational
modes of  WDs (Metcalfe et al. 2001). These uncertainties will affect the efficiency to separate the
 contribution of  $M_{MS}$ and Z, respectively, i.e. the reason for $C/O_{Mch}$ but not the observable relations
for LCs and spectra.

 We provide limits on the size of evolutionary effects due to $M_{MS}$ and metallicity of the
progenitor ($\Delta M \approx 0.2^m$). Other evolutionary effects may be due to a systematic
 change in the dominant  progenitor
scenario (e.g. mergers vs. single degenerate) or the
 typical separation in binary systems which contribute to the SNe~Ia at a given time.
The latter may alter the accretion rates and, consequently, the central densities of the WD at 
the time of the thermonuclear runaway.

 We did not consider the effect of the progenitor and of its pre-conditioning 
 just prior to the explosion
on the propagation of the burning front and, in particular, on the deflagration to
detonation transition or, alternatively, the  phase of transition from a slow to a 
 very fast deflagration (Hillebrandt, 1999 private communication).
 Although the  model parameters
have been chosen to allow for a representation of  ``typical" SNe~Ia,
more comprehensive studies and detailed fitting of actual observations
are needed   e.g. to detangle effects due to a change in the ignition density vs. the progentor.
 In particular, observations of local SNe~Ia have to be employed
to narrow down and test for the proposed range of flux ratio between  maximum  and tail,  and its
relation to the expansion velocities.

\bigskip

\noindent {\bf Acknowledgments:}
This work has been supported  by  NASA Grant NAG5-7937, the
 MURST italian grant Cofin2000,
by the MEC spanish grant PB96-1428, by the Andalusian grant FQM-108
 and it is part of the ITALY-SPAIN
integrated action (MURST-MEC agreement) HI1998. The calculations for the explosion and 
light curves were done on a Beowulf-cluster financed by the John W. Cox-Fund
of the Department of Astronomy at the University of Texas.

{} 
\vfill

{\bf        }
\begin{deluxetable}{llcccccc}
\tablenum{1}
\tablewidth{0pt}
\tablecaption{Properties of the models}
\tablehead { \colhead{} &
\colhead{Model} &
\colhead {M$_{MS}(M_\odot)$} &
\colhead {M$_{CO}^{TP}(M_\odot)$} &
\colhead {C$_{cen}$} &
\colhead {M$_{cen}(M_\odot)$} &
\colhead {C/O$_{Mch}$} &
\colhead {$^{56}Ni$ $(M_\odot)$}} 
\startdata
Z=0.02 & 1p5z22  & 1.5  & 0.55 & 0.21 & 0.27 &  0.75 & 0.589 \nl
Y=0.28  & 3p0z22 & 3.0  & 0.57 & 0.21 & 0.28 &  0.76  & 0.584 \nl
        & 5p0z22 & 5.0  & 0.87 & 0.29 & 0.46 &  0.72 & 0.561 \nl
      & 7p0z22   & 7.0  & 0.99 & 0.28 & 0.70 &  0.60 & 0.516 \nl
            &       &     &      &      &      &      &             \nl
Z=10$^{-3}$ & 1p5z13 & 1.5  & 0.59 & 0.24 & 0.31 & 0.76 & 0.587 \nl
Y=0.23      & 3p0z13 & 3.0  & 0.77 & 0.26 & 0.39 &  0.74 & 0.567 \nl
Y=0.23      & 5p0z13 & 5.0  & 0.90 & 0.29 & 0.58 &  0.66 & 0.541 \nl
            & 6p0z13 & 6.0  & 0.98 & 0.29 & 0.71 &  0.60 & 0.522 \nl
            &      &     &      &      &      &      &   \nl
            & LOW Rate     &     &      &      &      &      &   \nl
           & 3p0z13LR & 3.0 & 0.76 & 0.51 & 0.38 & 1.22 & 0.620 \nl
            &      &      &      &      &      &      &         \nl
Z=10$^{-4}$ & 3p0z14 & 3.0  & 0.80 & 0.27 & 0.41 & 0.73 & 0.568\nl
Y=0.23      & 5p0z14 & 5.0  & 0.90 & 0.29 & 0.58 & 0.65 & 0.541\nl
            & 6p0z14 &6.0   & 0.99 & 0.28 & 0.72 & 0.59 & 0.511 \nl
            &      &      &      &      &      &      &   \nl
Z=10$^{-10}$& 5p0z00 & 5.0  & 0.89 & 0.32 & 0.49 & 0.70 & 0.549 \nl
Y=0.23      & 7p0z00 & 7.0  & 0.99 & 0.31 & 0.59 & 0.62 & 0.525 \nl
\enddata
\end{deluxetable}
\begin{deluxetable}{ccccccc}
\tablenum{2}
\tablewidth{0pt}
\tablecaption{Element production (in $M_\odot$) of the reference model $5p0z22$.}
\tablehead{}
\startdata
 He & C & O & Ne & Na & Mg & Si \nl
6.62E-04 & 1.19E-02 & 9.18E-02 & 5.34E-03 & 4.91E-05 & 1.82E-02 &  2.61E-01 \nl 
 & & & & & & \nl
 P & S & Cl & Ar & K & Ca & Sc \nl
2.51E-05 & 1.59E-01 & 4.00E-06 & 3.28E-02 & 1.99E-06 & 3.44E-02 & 1.02E-08 \nl  
 & & & & & & \nl
Ti & V & Cr & Mn & Fe & Co & Ni \nl
1.61E-05 & 9.07E-04 & 6.56E-04 & 2.68E-02 & 6.57E-01 & 6.15E-03 & 6.47E-02 \nl 
\enddata
\end{deluxetable}

\vfill
\newpage
{\bf }
{
\begin{figure}
\caption{
 Final chemical Carbon (solid) and Oxygen (dotted) profiles
in the central region of
stars between 1.5 and 7 $M_\odot$ for solar abundances Z = 0.02. 
}
\label{z22}
\end{figure}
\begin{figure}
\caption{
Same as Fig. \ref{z22} but for stars with 3 $M_\odot$ and metallicities Z between
$10^{-10}$  and 0.02. 
}
\label{m3}
\end{figure}
\begin{figure}
\caption{
Same as Fig. \ref{z22} but for stars with 5 $M_\odot$ and metallicities Z between
$10^{-10}$  and 0.02. 
}
\label{m5}
\end{figure}
\begin{figure}
\vskip 10.cm
\caption{
 Influence of the nuclear reaction rate of $^{12}C(\alpha ,\gamma )^{16}O$
 on the final chemical profiles
of C(solid) and O(dotted) for a star with $3M_\odot$ and Z=0.001.
  The high and low rates are taken from Caughlan et al. (1985) and  
Caughlan \& Fowler (1988), respectively.
}
\label{CORate}
\end{figure}
\begin{figure}
\caption{
Final density (left scale) relative to $\rho$ at the center   and velocity profiles (right scale in
1000 km/sec) of the delayed detonation models.
The results are given for progenitors with
 $M_{MS}= 5 M_\odot$ with $Z=10^{-10},0.02$ (5p0z22+5p0z00,
left panel), and for progenitors with  $M_{MS}$= 1.5 \&
7 $M_\odot$ with Z=0.02 (1p5z22+7p0z22, right panel).
The velocity and density of model $1p5z22$ correspond to the higher and lower function, respectively.
 For the same $M_{MS}$ but different Z, the curves are indistinguishable.
}
\label{rho}
\end{figure}
\begin{figure}
\caption{
Final chemical profiles for delayed detonation models of progenitors with
$MS$ masses between 1.5 and 7 $M_\odot$.          
}
\label{chem}
\end{figure}
\begin{figure}
\caption{
Isotopic abundances relative to solar for models with  progenitors of various
$MS$ masses and metallicities. Isotopes of the same element are connected by lines.
}
\label{isotope}
\end{figure}
\begin{figure}
\caption{
B and V light curves for DD-models with progenitors of various $M_{MS}$ (upper panel)
 and metallicities (middle panel). In the lower panel, we show
the influence of the nuclear reaction rates  according
to Caughlan et al. (1985, 3p0z13) and Caughlan \& Fowler (1988, 3p0z13LR).
}
\label{lc}
\end{figure}
\begin{figure}
\caption{
 Influence of  $M_{MS}$ (left) and metallicity (right) on B (---), V ($^{....}$) and
 B-V (- - -) at maximum light. All quantities are given relative to the reference model $5p0z22$.
 In the right panel, the numbers 1,2,3,4 refer to Z of 0.02, 0.001, 0.0001
and $10^{-10}$, respectively.
}
\label{dep}
\end{figure}
\begin{figure}
\caption{
Fig. 10: {$\gamma $-luminosity normalized  to the instantaneous energy production by radioactive decays (in \%) as
a function of time for the reference model 5p0z22 (upper left). For other models, the difference 
is shown relative to model 5p0z22.}
}
\label{gam}
\end{figure}
}
\end{document}